# Die Einweg-Lichtgeschwindigkeit auf der rotierenden Erde und die Definition des Meters

*Peter Ostermann*

Seit 1983 ist das Meter definiert als *„Länge der Strecke, die Licht im Vakuum während der Zeit von 1/299792458 s durchläuft"* [1]. Gäbe es genau ein einziges widerspruchsfrei durchführbares Synchronisationsverfahren für natürliche Uhren, oder wären alle entsprechenden Verfahren äquivalent, so könnte man aus der Gültigkeit der speziellen Relativitätstheorie auf einen eindeutigen, vom Synchronisationsverfahren unabhängigen Wert der Einweg-Lichtgeschwindigkeit $c$ in Inertialsystemen schließen. Es ist diese offenbar allgemein akzeptierte Auffassung einer unbedingten Konstanz der Lichtgeschwindigkeit in Inertialsystemen, die der zitierten Meter-Definition zugrunde liegt.

Zwar ist es richtig, daß hinreichend langsam auseinandergeschobene Uhren beim Austausch von Lichtsignalen immer Reflexion im Zeitmittelpunkt anzeigen. Eine einfache Überlegung zeigt jedoch die − im Prinzip mit *einer einzigen* Uhr, d.h. unabhängig von jeder Synchronisation − nachweisbare Orts- und Richtungsabhängigkeit der Einweg-Lichtgeschwindigkeit in rotierenden Systemen, aus der sich zwangsläufig eine Verletzung der unbedingten Konstanz der Lichtgeschwindigkeit in lokalen Inertialsystemen ergibt. Es wird gezeigt, daß dies keinen Widerspruch zur ursprünglichen EINSTEIN'schen Relativitätstheorie darstellt, wohl aber zur aktuellen Meter-Definition.

Mit Berücksichtigung einer einfachen, hier formulierten Synchronisations-Bedingung ist eine globale *systeminterne* Synchronisation ortsfester Uhren auf der Erde prinzipiell möglich, und zwar dadurch, daß ein entsprechendes Verfahren der unterschiedlichen Einweg-Lichtgeschwindigkeit in stationären Systemen Rechnung trägt. Dagegen ist eine solche Synchronisation mit dem EINSTEINschen Prinzip der Reflexion im Zeitmittelpunkt bekanntlich nicht möglich.

Natürliche Maßstäbe und Uhren zeigen nicht den wahren Raum oder die wahre Zeit. Im Rahmen der *allgemeinen* Relativitätstheorie verlangt die denkbar einfachste Behandlung der rotierenden Scheibe keine lokale LORENTZ-, sondern eine GALILEI-Transformation. Im Unterschied zur herkömmlichen Interpretation und im Hinblick auf das längst konkret bestimmte kosmische Ruhsystem lassen sich die unverzichtbaren Systemkoordinaten der allgemeinen Relativitätstheorie sehr einfach verstehen als wahre Repräsentanten des absoluten euklidischen Raums und der absoluten kosmischen Zeit.

Sowohl die Versuche von SAGNAC bzw. MICHELSON und GALE – auf denen auch sehr aktuelle Experimente mit Laserkreiseln beruhen – als auch das Experiment von HAFELE und KEATING werden bisher mit Bezug auf ein übergeordnetes Inertialsystem erklärt. *Intern* können sie nur mit Berücksichtigung der orts- und richtungsabhängigen Einweg-Lichtgeschwindigkeit in stationären Systemen verstanden werden.

Angesichts der demzufolge prinzipiell meßbaren Abweichungen der Einweg-Lichtgeschwindigkeit auf der rotierenden Erde wird eine Modifizierung der Definition des Meters auf Basis des unbedingt konstanten *Durchschnittswerts c* der Lichtgeschwindigkeit für Hin- und Rückläufe vorgeschlagen, die von der Einstellung der Zeit-Nullpunkte verschiedener Uhren unabhängig ist.

Anhang: Nach EINSTEINs allgemein akzeptierter Auffassung sollte es sich bei Längenkontraktion und Zeitdilatation um rein kinematische Effekte handeln, die – im Unterschied zur Auffassung von LORENTZ und POINCARÉ – keiner dynamischen Erklärung bedürfen. Am EHRENFEST'schen Paradoxon der rotierenden Scheibe aber wird gezeigt, daß eine *scharfe* Trennung von relativistischer Kinematik und Dynamik prinzipiell nicht möglich ist. Daß es solch eine Einschränkung geben muß, ist zwar längst bekannt – bisher allerdings nur aus der Quantenmechanik.

---

# Inhalt

## 1. Nachweis der Richtungsabhängigkeit

Der Nachweis, daß die Einweg-Lichtgeschwindigkeit in rotierenden Systemen *nicht* konstant sein kann, ist unabhängig vom Ausgang irgendeines realen Experiments – unter der einzigen Voraussetzung, daß sie in demjenigen System konstant ist, in welchem die Drehachse ruht.

Wir betrachten einen nicht notwendigerweise starren, aber stationär rotierenden Kreisring bzw. eine entsprechende Kreisscheibe S*, auf der eine Lichtquelle L* mit Uhr sowie ein Satz von *n* gleichmäßig verteilten Spiegeln $S_1^*$, $S_2^*$ ... $S_n^*$ fest installiert sind (s. Abb. 1). Die Drehachse von S* ruht in einem Inertialsystem S, in dem die Lichtgeschwindigkeit nach Voraussetzung den konstanten Betrag *c* hat. Bei Verwendung hinreichend vieler Spiegel[1]) können kreisförmige Umläufe von Lichtsignalen in einer für alle praktischen Belange genügenden Annäherung realisiert werden. Von zwei Lichtsignalen, die in entgegengesetzter Richtung gleichzeitig von L* ausgehen und den Ring S* auf angenäherten Kreisbahnen umlaufen, kehrt bekanntlich dasjenige früher zu L* zurück, das entgegen der Drehrichtung von S* gelaufen ist. Diese Feststellung steht außer Zweifel, da das eine Lichtsignal wegen der Drehgeschwindigkeit $v = \omega r$ der Lichtquelle L* im Inertialsystem S einen kürzeren Weg zurückgelegt hat als das andere. Für den Grenzfall kreisförmiger Bahnen beträgt der entsprechende Laufzeitunterschied:

$$\Delta T = \frac{4\pi r v / c^2}{1 - v^2 / c^2}. \tag{1}$$

Einen von Null verschiedenen Laufzeitunterschied gibt es aber auch im rotierenden System S*. Und zwar gilt diese Aussage unabhängig von Ganggeschwindigkeit und Zeitnullpunkt einer mitbewegten Uhr bei L*. Denn wie schnell diese Uhr im Vergleich zu denen des Systems S auch geht – wenn sie überhaupt nur irgendwie vorwärts geht, dann muß sie für die Rückkehr der beiden gegenläufigen Lichtsignale ebenfalls *unterschiedliche Zeitpunkte* anzeigen:

$$\Delta T^* \neq 0.$$

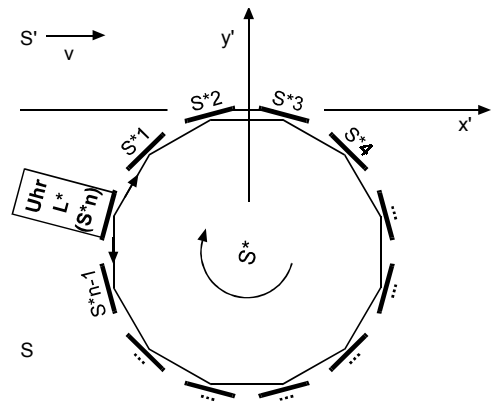

**Abb. 1:** Weil die Lichtgeschwindigkeit in S konstant ist, kehren zwei gegenläufige, von L* gleichzeitig emittierte Lichtsignale auch zu verschiedenen Zeiten t* zur *mitbewegten* Uhr in S* zurück - ganz unabhängig von jeder Synchronisation.

Das läßt aber gar keine andere Deutung zu, als daß die Geschwindigkeiten der beiden Lichtsignale, die ja im rotierenden System S* den gleichen Weg[2]) zurückgelegt haben, verschieden sein müssen: *Der Betrag der Einweg-Lichtgeschwindigkeit in rotierenden Systemen ist abhängig von Ort und Richtung.*

Weil schließlich aber zu dieser bloßen Feststellung unterschiedlicher Rückkehrzeiten bei zwei gleichzeitig aus der gleichen Quelle ausgesandten Lichtsignalen die Verwendung *einer einzigen Uhr ganz beliebiger Ganggeschwindigkeit* genügt, ist die hier nachgewiesene Abhängigkeit der Einweg-Lichtgeschwindigkeit von Ort und Richtung in rotierenden Systemen als qualitative Feststellung vollständig unabhängig von allen diesbezüglichen Erkenntnissen der speziellen und der allgemeinen Relativitätstheorie.

## 2. Das Paradoxon der Einweg-Lichtgeschwindigkeit

Aus dieser Orts- und Richtungsabhängigkeit der Lichtgeschwindigkeit in rotierenden Systemen scheint sich nun aber ein Widerspruch in den Grundlagen der EINSTEIN'schen speziellen Relativitätstheorie zu ergeben:

---

[1]) EINSTEIN selbst hat von dieser Voraussetzung der polygonalen Approximation bereits in seiner grundlegenden Arbeit von 1905 Gebrauch gemacht: *„Nimmt man an, daß das für eine polygonale Linie bewiesene Resultat auch für eine stetig gekrümmte Kurve gelte, so erhält man den Satz ..."* – A. EINSTEIN [2], S. 904

[2]) Dies gilt für den Grenzfall kreisförmiger Bahnen mit Mittelpunkt auf der Drehachse. Im Unterschied zu diesen sind beliebige polygonale Lichtwege für wechselnde Laufrichtung nicht genau identisch. V. LAUE [3], Bd. I, weist aber darauf hin, daß die resultierenden Unterschiede nach dem FERMAT'schen Prinzip klein sind in $O(v^2/c^2)$ und damit ohne praktische Bedeutung für die Laufzeitunterschiede, diese selbst sind klein in $O(v/c)$.

**I.** Ist die Lichtgeschwindigkeit in einem Inertialsystem S konstant, so ist sie bezüglich einer in S rotierenden Scheibe S* *nicht* konstant. Ihr Betrag unterscheidet sich auf S* erstens mit der Entfernung von der Drehachse und zweitens an ein und demselben Ort von S* je nach Ausbreitungsrichtung.

**II.** Da die Lichtgeschwindigkeit demzufolge auf einem beliebig herausgegriffenen infinitesimal kleinen Stück der rotierenden Scheibe je nach Laufrichtung verschieden ist, so ist sie es auch in einem entsprechend bewegten Inertialsystem, in dem dieses Stück während der infinitesimal kurzen Laufzeiten der beiden Signale in hinreichender Näherung ruht.

**III.** Dies steht in klarem Widerspruch zur offenbar allgemein akzeptierten Auffassung[3]), daß nämlich die EINSTEIN'sche spezielle Relativitätstheorie den konstanten Wert *c* der Einweg-Geschwindigkeit von Lichtsignalen in Inertialsystemen impliziert.

Hier ist von entscheidender Bedeutung, daß die oben festgestellte Orts- und Richtungsabhängigkeit der Lichtgeschwindigkeit wegen der vorliegenden Symmetrieverhältnisse auch für beliebig kleine Teilabschnitte der durchlaufenen Bahn, d.h. nicht nur global sondern auch *lokal* gelten muß. Denn wenn auch die Differenz der Laufzeitintervalle zweier in entgegengesetzter Richtung umlaufender Signale für ein und dieselbe – immer kleiner werdende – infinitesimale Teilstrecke gegen Null geht, so bleibt doch ihr Quotient verschieden von Eins. Dieser Quotient ist aber nichts anderes als das Verhältnis der beiden unterschiedlichen Werte der tangentialen Einweg-Lichtgeschwindigkeit auf der entsprechenden Kreisbahn des rotierenden Systems S*. Es ist dabei sehr bemerkenswert, daß für ein und dieselbe vorgegebene Geschwindigkeit *v* des Scheibenrands die Abweichung eines entsprechend kleinen Bereichs der Peripherie von einem idealen lokalen Inertialsystem S' mit zunehmendem Scheibenradius *in jeder gewünschten Annäherung schwindet.*

## 3. Elementare Berechnung für tangentiale Ausbreitung

Die Relativitätstheorie, und zwar zunächst die spezielle, kommt erst jetzt ins Spiel, wenn nach den konkreten Werten der Einweg-Lichtgeschwindigkeit gefragt wird. Wir bleiben beim aufschlußreichen Grenzfall kreisförmiger Lichtwege um die Drehachse, was tangentialen Lichtwegen auf hinreichend kleinen Teilstrecken entspricht. Weil die Laufzeiten $T_\pm^*$ von einer einzigen an der Rotation teilnehmenden Uhr bei L* abgelesen werden, ergeben sich diese mit Berücksichtigung der Zeitdilatation aus den Laufzeiten $T_\pm$ des Inertialsystems $S$ zu

$$T_\pm^* = T_\pm \sqrt{1-\frac{v^2}{c^2}} = \frac{2\pi r \sqrt{1-\frac{v^2}{c^2}}}{c \mp v}. \tag{2}$$

Weiterhin gilt für den von einem mitbewegten Beobachter gemessenen Umfang *U\** der rotierenden Scheibe S* gegenüber dem Umfang *U* eines deckungsgleichen Kreises im Inertialsystem S wegen der FITZGERALD-LORENTZ-Kontraktion bekanntlich[4]):

$$U^* = \frac{U}{\sqrt{1-\frac{v^2}{c^2}}} = \frac{2\pi r}{\sqrt{1-\frac{v^2}{c^2}}}. \tag{3}$$

Ohne jeden Vorgriff auf den mathematischen Apparat der allgemeinen Relativitätstheorie, ergeben sich allein daraus die beiden gesuchten Werte der Einweg-Lichtgeschwindigkeit für entgegengesetzte Laufrichtungen im rotierenden System S* zu

$$c_\pm = \frac{U^*}{T_\mp^*} = \frac{c}{1 \mp \frac{v}{c}}. \tag{4}$$

Bei klassischer Berechnung hätte man stattdessen die beiden Werte $c \pm v$ erhalten. Diese stimmen zwar mit den soeben gefundenen in erster Ordnung überein, weisen aber bei näherem Hinsehen einen wesentlichen Unterschied auf. Der einfache Ausdruck (4) zeigt nämlich die bemerkenswerte Eigentümlichkeit, daß er einerseits der unterschiedlichen Einweg-Lichtgeschwindigkeit Rechnung trägt, andererseits aber bei Hin- und Rückläufen auf beliebigen Teilstrecken der hier behandelten Kreisbahn für die Durchschnittsgeschwindigkeit der Lichtsignale immer den exakten Wert $c$ liefert.

---

[3]) wie sie eben in der aktuellen Meterdefinition deutlich zum Ausdruck kommt

[4]) Auf das daraus resultierende EHRENFEST'sche Paradoxon [4] werden wir im Anhang dieser Arbeit ausführlich eingehen.

**Prinzip der lokalen Durchschnittsgeschwindigkeit $c$ für hin- und zurücklaufende Lichtsignale:** *In jedem Inertialsystem ist der mit natürlichen Maßstäben und Uhren gemessene Durchschnittswert der Lichtgeschwindigkeit für Hin- und Rückläufe auf demselben Weg gleich der Naturkonstanten $c$.*

Dies ist deshalb von weitreichender Bedeutung, weil sich Längenkontraktion, Zeitdilatation und selbst eine allen Erfahrungstatsachen genügende Theorie der Inertialsysteme – auch *ohne* Relativität der Gleichzeitigkeit – allein aus dem Prinzip der konstanten Durchschnittsgeschwindigkeit $c$ in Verbindung mit dem EINSTEIN'schen Relativitätsprinzip in seiner ursprünglichen Formulierung ableiten lassen. Und genau darin liegt auch die eigentliche Rechtfertigung für die Anwendung der Formeln von Längenkontraktion und Zeitdilatation auf rotierende Systeme, in denen zwar die Lichtgeschwindigkeit bezüglich ihres Durchschnittswertes konstant ist, nicht aber die Einweg-Geschwindigkeit von Lichtsignalen.

### 4. Die *bedingte* Konstanz der Einweg-Lichtgeschwindigkeit

Zur endgültigen Klärung des oben aufgezeigten Paradoxons denken wir uns an den Enden zweier Einheitsmaßstäbe, die sich für die Dauer des Experiments unmittelbar nebeneinanderher in Längsrichtung bewegen sollen, jeweils Uhren gleicher Ganggeschwindigkeit angebracht. Eine Lichtwelle, deren Ausbreitungsrichtung parallel zur Ausrichtung der Maßstäbe liegt, möge zuerst die Uhren am jeweils linken Ende, eine kurze Zeit später die Uhren am jeweils rechten Ende der beiden Maßstäbe erreichen. Die entsprechenden Zeitpunkte werden registriert, woraus sich die Laufzeiten ergeben:

a) Die Uhren an den Enden des einen Maßstabs bewegen sich beide mit konstanter Geschwindigkeit $v$ exakt auf einer Geraden. Wir befinden uns in einem Inertialsystem. Die Lichtgeschwindigkeit, die sich aus den registrierten Zeitpunkten dieser beiden Uhren ergibt, sei gleich $c$.

b) Die Uhren an den Enden des anderen Maßstabs bewegen sich beide mit konstanter Geschwindigkeit $v = \omega r$ auf dem Bogen eines Kreises, der so groß ist, daß sich auch bei höchster Präzision der Messung keinerlei Abweichungen im Vergleich zu der geradlinig-gleichförmigen Bewegung der erstgenannten Uhren feststellen lassen. Die Lichtgeschwindigkeit, die sich aus den registrierten Zeitpunkten dieses zweiten Uhrenpaares ergibt, ist aufgrund des oben gefundenen Ausdrucks (4) gleich $c/(1\pm \omega r/c)$, was unter der Bedingung $\omega r < c$ je nach Radius $r$ einem Wert zwischen $c/2$ und $\infty$ entsprechen kann.

Dasselbe Lichtsignal bewegt sich also mit vollkommen *verschiedener* Geschwindigkeit gleichzeitig über zwei, während der Dauer des Experiments relativ zueinander in Ruhe befindliche, unmittelbar nebeneinander liegende Einheitsmaßstäbe hinweg, *wobei alle vier verwendeten Uhren gleich schnell gehen*. Wie löst sich dieser Widerspruch? – Es ist kein Widerspruch.

Dies deshalb, weil aufgrund unterschiedlich eingestellter Zeitnullpunkte beides nebeneinander möglich ist. Im Falle a) sind die Zeitnullpunkte der Uhren *lokal* synchronisiert nach dem EINSTEIN'schen Prinzip, im Falle b) jedoch *global* nach einem allgemeineren Synchronisations-Prinzip, das wir im folgenden auffinden wollen.

Der scheinbare Widerspruch läßt sich also durch bloßes Verstellen der Zeitnullpunkte auflösen: *Im Unterschied zur Durchschnittsgeschwindigkeit $c$ ist die Einweg-Lichtgeschwindigkeit in Inertialsystemen ohne Angabe des gewählten Synchronisationsverfahrens nicht eindeutig festgelegt, sondern systemintern eine unbestimmte Größe.*

### 5. EINSTEINs ursprüngliche Formulierung

Daraus folgt, daß das Prinzip einer unbedingten Konstanz der Lichtgeschwindigkeit, wie es z.B. der aktuellen Meterdefinition zugrunde liegt, nicht aufrecht zu erhalten ist. Zwar ist es offenbar richtig, daß die Lichtgeschwindigkeit in einem beliebigen Inertialsystem bei geeigneter Synchronisation konstant gleich $c$ sein *kann*. Aber es ist falsch, daß sie gleich $c$ sein *muß*. Und dies steht ganz im Einklang mit EINSTEINs ursprünglicher Formulierung:

„Die letztere Zeit kann nun definiert werden, indem man *durch Definition* festsetzt, daß die ‚Zeit', welche das Licht braucht, um von A nach B zu gelangen, gleich ist der ‚Zeit', welche es braucht, um von B nach A zu gelan-gen ... Wir nehmen an, daß diese Definition des Synchronismus in widerspruchsfreier Weise möglich sei ..." *(Hervorhebung wie im Original) – [2], S. 894*

EINSTEINS ‚*Definition*' ist also in allen Inertialsystemen widerspruchsfrei möglich – aber sie ist nicht zwingend. Die spezielle Relativitätstheorie setzt lediglich voraus, daß sich der universelle Wert $c$ einer richtungsunabhängigen Einweg-Lichtgeschwindigkeit durch eine entsprechende Synchronisation der Uhren in jedem Inertialsystem realisieren läßt.

Das von EINSTEIN dazu begründete Synchronisationsverfahren der Reflexion im Zeitmittelpunkt aber zeichnet sich gegenüber anderen denkbaren Verfahren dadurch aus, daß es in allen Inertialsystemen zur Synchronisation durch langsamen Uhrtransport äquivalent ist. Eine hinreichend langsam verschobene Uhr bleibt hier also im Sinne der speziellen Relativitätstheorie synchronisiert. Allerdings führt dieses Verfahren bereits für einfachste Nicht-Inertialsysteme zu Widersprüchen. Nach dem EINSTEIN'schen Prinzip wäre insbesondere eine *systeminterne* Synchronisation der Uhren auf der Erde prinzipiell unmöglich, selbst wenn man diese als ideales Geoid[5]) betrachtet, auf welchem alle Uhren gleich schnell gehen.

Bereits KALUZA [5] hat auf „die theoretische Möglichkeit eines Nachweises der Erdrotation durch rein optische bzw. elektromagnetische Experimente" hingewiesen[6]), wobei er einen ‚Schlußfehler' der Synchronisation einführt und diesen mit maximal $2*10^{-7}$ s beziffert.

Aus unserer Sicht entspricht dies gerade der Differenz der mit einer einzigen Uhr meßbaren Laufzeit eines elektromagnetischen Zeitsignals um den Äquator, verglichen mit der Laufzeit, die sich errechnet, wenn man fälschlicherweise eine Konstanz der Einweg-Lichtgeschwindigkeit unterstellt[7]).

Daß aber zumindest im Spezialfall eines rotierenden Ringes sogar eine *natürliche* Synchronisation – ohne jeden technischen Eingriff in die Ganggeschwindigkeit mitbewegter natürlicher Uhren – möglich sein muß, ergibt sich aus folgender Überlegung: Angenommen der Ring S* befindet sich zunächst ruhend in einem übergeordneten Inertialsystems S. Alle Uhren sollen hier synchronisiert sein. Wird nun S* allmählich in gleichmäßige Rotation versetzt, so kann die Synchronisation der Uhren von S* schon aus Symmetriegründen nicht aufgehoben werden, weil sich in diesem speziellen Fall eine mögliche rotationsabhängige Veränderung der Ganggeschwindigkeit auf alle Uhren gleichmäßig auswirken muß.

Doch diese Voraussetzung ist bereits auf einer rotierenden Scheibe nicht mehr erfüllt, ein solches Verfahren kommt für die Erde ohnehin nicht in Betracht. Ist es nicht trotzdem auch in diesen Fällen möglich, die EINSTEIN'sche Synchronisationsvorschrift durch eine allgemeinere, widerspruchsfreie aber ebenfalls systeminterne zu ersetzen?

## 6. Allgemeine Synchronisations-Bedingung

Was immer ein Synchronisationsverfahren auch leisten mag, eines ist ganz unverzichtbar – am Ende hat jede Uhr notwendigerweise synchron zu gehen zu sich selbst. Ist dies nicht der Fall, so kann man die These vertreten, dies liege daran, daß eine vollständige Synchronisation innerhalb des betreffenden Systems prinzipiell nicht widerspruchsfrei möglich sei. Vorher jedoch stellt sich die Frage, ob das gewählte Synchronisationsverfahren nicht einfach auf falschen Voraussetzungen beruht.

Zunächst einmal ist klar, daß jede vernünftige Synchronisation[8]) die folgende, unmittelbar aus dem Kausalitätsprinzip fließende Bedingung erfüllen muß:

1.) Allgemeine Synchronisations-Bedingung: *Uhren sind nur dann richtig synchronisiert, wenn alle durch das Vakuum übertragenen elektromagnetischen Zeitsignale einer beliebigen Uhr bei jeder an-*

---

[5]) s. Abschnitt 7

[6]) s. Abschnitt d) und e) des Anhangs

[7]) Seit damals wird eine unbedingte Konstanz der Lichtgeschwindigkeit in rotierenden Systemen gewissermaßen dadurch gewährleistet, daß man erstens die Uhren auf jedem einzelnen Lichtweg so synchronisiert, daß die Einweg-Lichtgeschwindigkeit per definitionem gleich der Konstanten $c$ ist, zweitens deklariert, daß dies nur entlang nichtgeschlossener Lichtwege möglich ist, drittens aber für geschlossene Wege einem ‚Schlußfehler' einführt, der betragsmäßig gerade die falsche Voraussetzung einer konstanten Lichtgeschwindigkeit ‚erklärt'.

[8]) Unter Synchronisation verstehen wir i. a. die Anpassung sowohl der *Zeitnullpunkte* als auch der *Ganggeschwindigkeiten* technischer Uhren.

*deren Uhr in Zeitpunkten reflektiert werden, die zwischen den Zeitpunkten ihrer Aussendung und Rückkehr liegen.*

Wir betrachten nun verschiedene Typen stationärer Systeme: a) die stationäre Translation, b) die stationäre Rotation, c) das stationär schwingende System. Die Stationarität ist hier gegebenenfalls auch statistisch[9]) zu verstehen, wodurch auch Vielteilchensysteme wie z.B. eine stationäre Atmosphäre (a), Luftwirbel (b) und stehende Schallwellen (c) eingeschlossen sind (das statische System ist jeweils als Grenzfall enthalten).

Alle stationären Systeme haben gewisse räumlich-stationäre Punkte, zwischen denen die zunächst mit *natürlichen Uhren* gemessenen Zeitspannen für hin- und zurücklaufende elektromagnetische Signale immer die gleichen bleiben. Denkt man sich nun aber alle Zeitpunkte an *technischen Systemuhren* abgelesen, die sich an Ort und Stelle des jeweiligen Ereignisses befinden und bezüglich Zeitnullpunkt und Ganggeschwindigkeit einstellbar sind, dann gilt die folgende

2.) Bedingung für stationäre Systeme: *Seien in allen räumlich-stationären Punkten eines abgeschlossenen Systems technische Uhren angebracht, so haben diese nur dann die richtige Ganggeschwindigkeit, wenn die durch das Vakuum übertragenen elektromagnetischen Zeitsignale einer beliebigen Uhr bei jeder anderen Uhr – gegebenenfalls statistisch gemittelt – immer mit derselben, für dieses Uhrenpaar charakteristischen Verzögerung eintreffen.*

Die allgemeine Bedingung 1.) gewährleistet eine dem Kausalitätsprinzip entsprechende zeitliche Reihenfolge und damit die Vermeidung irgendwelcher `Schlußfehler`. Die Bedingung 2.) stellt – i. a. durch technische Eingriffe – die gleiche Ganggeschwindigkeit aller Systemuhren sicher. Beide Bedingungen zusammengenommen ermöglichen bereits eine widerspruchsfreie, allerdings noch keine eindeutige Synchronisation. Doch eine solche kann es für abgeschlossene *Teilsysteme* auch gar nicht geben[10]), wenn sich diese nämlich nach Voraussetzung ohne Bezug auf das übergeordnete System physikalisch vollständig beschreiben lassen.

Es ist nun aber bemerkenswert, daß sich die denkbar einfachste systeminterne Synchronisation im Spezialfall eines rotierenden Rings Schritt für Schritt durchführen läßt, ohne dabei auf ein externes, übergeordnetes Inertialsystem Bezug nehmen zu müssen, und zwar bei Verwendung der folgenden

3.) Zusatzbedingung für symmetrische Systeme: *Sind die Uhren eines symmetrischen räumlich-stationären Systems richtig synchronisiert, so sind – bei Beachtung der Ausbreitungsrichtung – die Laufzeiten für alle Lichtwege gleich, deren Endpunkte durch symmetriebezogene Operationen aufeinander abgebildet werden können. Symmetriebezogene Operationen sind insbesondere die Parallelverschiebung in Inertialsystemen und die Drehung um die Achse in rotierenden Systemen.*

Diese Zusatzbedingung 3.) fließt ebenso wie 1.) aus dem Kausalitätsprinzip. Zusammen sorgen nun beide für einen gemeinsamen Zeitnullpunkt aller Uhren auf einem rotierenden Ring. Nach denselben Bedingungen könnte dann aber auch – und zwar systemintern – eine alternative Synchronisation beliebiger Inertialsysteme vonstatten gehen[11]). Insbesondere läßt sich in allen lokalen Inertialsystemen des Geoids (s. Abschn. 7) oder des rotierenden Rings $S^*$ jeweils sogar eine solche widerspruchsfreie Synchronisation der *natürlichen* Uhren durchführen, daß diese mit der globalen Synchronisation des gesamten stationären Systems $S^*$ bis auf einen gemeinsamen konstanten Dilatationsfaktor übereinstimmt.

Um nun aber zugleich in allen rotierenden Systemen *und* Inertialsystemen nicht nur eine widerspruchsfreie, sondern auch eine eindeutige, systeminterne Synchronisation zu erreichen, stellen wir schließlich die

---

[9]) Die Erdatmosphäre ist im Idealfall ein Beispiel solch eines statistisch-stationären Systems.

[10]) Auf den Sonderfall des einen und einzigen stationären offenen Systems wird in einer eigenen Arbeit d. Verf. „Ein stationäres Universum und die Grundlagen der Relativitätstheorie" ausführlich eingegangen, die voraussichtlich in Kürze hier erscheinen wird.

[11]) Dazu sind z.B. den Hin- und Rücklaufzeiten eines Lichtsignals vom Koordinatenursprung zu jedem beliebigen Punkt *(x',y',z')* des Inertialsystems S' die Werte

$$t'_{\pm} = \frac{1}{c}\sqrt{x'^2 + y'^2 + z'^2} \pm \frac{v\,x'}{c^2}$$

zuzuordnen, und zwar mit frei wählbarem *v/c* aus dem Intervall von -1 bis 1. Durch diese interne, widerspruchsfrei mögliche Synchronisation ist bei externer Betrachtung ein parallelachsiges Inertialsystem S als (lokales) Ruhsystem ausgezeichnet, gegen das sich das (lokale) Inertialsystem S' mit der Geschwindigkeit *v* in Richtung der positiven *x'*-Achse bewegt.

4.) Forderung der minimalen zeitlichen Abweichung: *Die Zeitnullpunkte hinreichend nah benachbarter, gleichmäßig verteilter Uhren sind, soweit dies nach Berücksichtigung der Punkte 1.) – 3.) noch möglich ist, nach dem Prinzip minimaler quadratischer Abweichung von der Reflexion im Zeitmittelpunkt einzustellen[12]).*

Die Schritte gemäß den *Bedingungen* 1.) – 3.) haben zuvor eine eindeutige Synchronisation aller Uhren eines jeden ringförmigen Teilbereichs bewirkt, sodaß jetzt nur noch eine für alle Uhren eines jeden einzelnen Rings gemeinsame Verstellung des Zeitnullpunktes erlaubt ist. Zur Erfüllung der abschließenden *Forderung* 4.) genügt es deshalb, diese mit Rücksicht auf die Symmetrieverhältnisse in S* nur auf solche benachbarten Uhren anzuwenden, die auf demselben Radius der rotierenden Scheibe bzw. in Nord-Süd-Richtung auf der rotierenden Erde liegen. Demgegenüber betrifft Forderung 4.) aber alle Uhren eines beliebigen Inertialsystems *S*.

Insgesamt ist nun also nach dem Verfahren gemäß 1.) – 4.) eine eindeutige, systeminterne Synchronisation der Uhren in allen stationären Systemen zumindest statistisch möglich. Das Verfahren führt auch im statischen Gravitationsfeld zu einem eindeutigen Ergebnis. Die Eindeutigkeit der *internen* Synchronisation kann sich allerdings immer nur auf das jeweils ins Auge gefaßte abgeschlossene stationäre bzw. statische System beziehen. So stimmt zwar die interne Synchronisation des rotierenden Systems Erde überein mit der externen des rotationsfreien Schwerpunktsystems Erde, nicht aber mit der des Teilsystems Erde bei Synchronisation des übergeordneten Sonnensystems.

Man könnte am Ende versuchen, den jetzt noch verbliebenen Spielraum zur systeminternen Wahl eindeutiger räumlicher Koordinaten zu nutzen, und zwar durch die abschließende

\*) Forderung der minimalen räumlichen Abweichung: *Zuletzt sind die Raum-Koordinaten, falls dies möglich ist, so zu wählen, daß der mit natürlichen Maßstäben gemessene räumliche Abstand solcher benachbarter Uhren, die beim Signalaustausch Reflexion im Zeitmittelpunkt zeigen, mit der Differenz einer entsprechenden Koordinate übereinstimmt .*

Diese Forderung \*) läßt sich im rotierenden System – als Spezialfall eines reinen Beschleunigungsfeldes – leicht, im wahren, statischen Gravitationsfeld aber gar nicht erfüllen. Das entspricht der Tatsache, daß im (lokalen) Gravitationsfeld eine eindeutige Zuordnung zwischen wahren Längen und den entsprechenden Koordinatendifferenzen nicht möglich ist: in solchen Teilsystemen – nicht aber im *kosmischen* Bezugssystem – gibt es kein eindeutiges Gravitationsgesetz, d.h. keine eindeutige Trennung von Gravitation und Geometrie. Im Unterschied zu den Bedingungen 1.) – 3.) handelt es sich bei 4.) und \*) ausdrücklich um Forderungen, die gegebenenfalls mittels einer verträglichen Koordinaten-Transformation durch andere ersetzt werden können.

## 7. Die globale Synchronisation auf der Erde

Eine globale *systeminterne* Synchronisation der Uhren auf der rotierenden Erde ist nicht nur möglich, sondern geradezu unverzichtbar[13]). Auf der Oberfläche des abgeplatteten Geoids gehen natürliche Uhren trotz breitengradabhängiger Rotationsgeschwindigkeit überall gleich schnell, weil sich Geschwindigkeitseffekt und Gravitationseffekt hier gerade aufheben. Dies ist von ALLEY und Mitarbeitern experimentell bestätigt [6]. *Auf einem idealen Geoid ist deshalb eine systeminterne globale Synchronisation sogar bei ausschließlicher Verwendung natürlicher Uhren möglich.*

Die aufgrund der Symmetrieverhältnisse einzig angemessene und zugleich denkbar einfachste systeminterne Synchronisation rotierender Systeme, die sich hier – im Unterschied zur EINSTEIN-Synchronisation – prinzipiell Schritt für Schritt gemäß den Bedingungen 1.) – 3.) sowie der Forderung 4.) bewerkstelligen läßt, führt im Ergebnis schließlich gerade auf die Systemzeit des übergeordneten Inertialsystems S zurück[14]). d.h. die systeminterne Synchronisation stimmt überein mit der üblichen, bisher alleine für möglich gehaltenen *externen* aus dem rotationsfreien Schwerpunktsystem. Die direkte Laufzeitmessung von Langwellensignalen auf der Erde mit global synchronisierten Uhren würde bei entsprechender technischer Qualität des Verfahrens demzufolge erweisen, daß die Einweg-Lichtgeschwindigkeit unterschiedlich ist, je nachdem, ob ein Signal mit oder entgegen der Erddrehung läuft. Möglicherweise genügt zu diesem Nachweis eine systeminterne Neubewertung bereits vorhandener Meßdaten.

---

[12]) Die nicht unproblematische, hier geforderte ‚gleichmäßige' Verteilung der Uhren ist mit Berücksichtigung der Symmetrieverhältnisse bei Bedarf näher zu spezifizieren.

[13]) Die Internationale Atomzeit TAI wird definiert als gewichteter Mittelwert aus über 200 Atomuhren weltweit. Es ist eine interessante Frage, ob sich die richtige Synchronisation stationärer Systeme möglicherweise allein aus den Bedingungen 1.) und 2.) in Kombination mit der Forderung 4.) *statistisch* ermitteln läßt.

[14]) Zunächst bis auf einen gemeinsamen konstanten Dilatationsfaktor, der sich z.B. aus dem Gravitationspotential an den abgeplatteten Polen ergibt, dann aber mit der Wahl einer geeigneten Zeiteinheit korrigiert werden kann.

### 8. Die Lichtgeschwindigkeit in der allgemeinen Relativitätstheorie

Die allgemeine Relativitätstheorie kennt von Anfang an verschiedene Werte für die Lichtgeschwindigkeit im Beschleunigungs- bzw. im Gravitationsfeld[15]. Dabei handelt es sich zunächst einmal um die mit natürlichen Maßstäben und EINSTEIN-synchronisierten, natürlichen Uhren zu messende ‚Lichtgeschwindigkeit' $c = konst$ des lokalen Inertialsystems. Demgegenüber stellt der ‚Koordinatenwert der Lichtgeschwindigkeit' $c^*_\pm$ eine auf die in Teilsystemen frei wählbaren Raum-Zeit-Koordinaten bezogene Lichtgeschwindigkeit dar. Nach dem oben gesagten ist nun aber auch die Berücksichtigung einer dritten Lichtgeschwindigkeit $c_\pm$ unumgänglich, die wir die *Einweg-Geschwindigkeit* für Lichtsignale in stationären Systemen nennen wollen, und bei der es sich um den auf ‚natürliche' Maßeinheiten bezogenen Wert der Einweg-Lichtgeschwindigkeit handelt. Aufgrund der Ergebnisse des Abschnitts 3 stimmt der *Durchschnittswert* dieser Einweg-Geschwindigkeit $c_\pm$ mit der Naturkonstanten $c$ überein. Schließlich definieren wir noch die *längenbezogene* sowie auch die *zeitbezogene* Einweg-Geschwindigkeit $c_{l\pm}$ bzw. $c_{\tau\pm}$ (s. Tab. 1). Setzt man im Linienelement des vierdimensionalen Raum-Zeit-Kontinuums, welches das Linienelement der speziellen Relativitätstheorie als Sonderfall enthält,

$$ds^2 = g_{ik}\, dx^i dx^k \tag{5}$$

für Lichtsignale $ds = 0$, und löst diese Gleichung nach $dx^0$ auf, so findet man zwei Lösungen, die der Ausbreitung in entgegengesetzte Richtungen entsprechen (im Unterschied zu lateinischen $i, k, l... = 0, 1, 2, 3$ zeigen griechische Indizes $\alpha, \beta, \gamma ... = 1, 2, 3$ nur die räumlichen Koordinaten an):

$$dx^0 = \frac{dl}{\sqrt{g_{00}}} - \frac{g_{0\alpha}\, dx^\alpha}{g_{00}} = \frac{dl}{\sqrt{g_{00}}} \pm \frac{g_{0\alpha}\left|dx^\alpha\right|}{g_{00}}. \tag{6}$$

Hier und im folgenden steht $dl$ für das bekannte Linienelement [7] der räumlichen Entfernung

$$dl^2 = \left(-g_{\alpha\beta} + \frac{g_{0\alpha}\, g_{0\beta}}{g_{00}}\right) dx^\alpha dx^\beta. \tag{7}$$

Dieses wurde seinerzeit für den Spezialfall der rotierenden Scheibe von KALUZA [5] aus der Voraussetzung gewonnen, daß die spezielle Relativitätstheorie im übergeordneten Inertialsystem gelten muß. Wir aber interpretieren es dahingehend, daß der gesuchte räumliche Abstand dem halben Produkt $c\, d\tau$ gleich sein muß, wobei $d\tau = dt\sqrt{g_{00}}$ das Intervall der Eigenzeit einer im Ausgangspunkt ruhenden Uhr[16] bedeutet, das unabhängig von der gewählten Synchronisation für Hin- und Rücklauf eines entsprechenden Lichtsignals insgesamt benötigt wird. Mit anderen Worten – zur Berechnung des räumlichen Linienelements bedient man sich seit langem einer Formel, die mit dem Prinzip von der Konstanz der *durchschnittlichen* Lichtgeschwindigkeit in Einklang steht.

Gehen wir einen einfachen Schritt weiter, indem wir voraussetzen, daß bei hinreichend kleinem Lichtweg $dl$ die Änderung von $g_{00}$ in der infinitesimalen Umgebung des gewählten Raumpunktes vernachlässigbar ist, ohne daß aber der Begriff der Geschwindigkeit zwischen zwei benachbarten Punkten seinen Sinn verliert. Mit der daraus resultierenden Erweiterung des üblichen Begriffs der Eigenzeit $d\tau$ auf eine infinitesimale Umgebung

$$dx^0 = c\, dt = \frac{c\, d\tau}{\sqrt{g_{00}}} \tag{8}$$

---

[15]) Unter Gravitationsfeld verstehen wir das *wahre* Gravitationsfeld ($R^i{}_{klm} \neq 0$), im Unterschied zum reinen Beschleunigungsfeld ($R^i{}_{klm} = 0$).

[16]) Die isolierten Bezeichnungen ‚Eigenzeit', ‚τ' oder ‚$d\tau$' beziehen sich immer auf die Eigenzeit einer ruhenden Uhr, bei einer bewegten Uhr soll es dagegen ausdrücklich ‚$s$' bzw. ‚$ds$' heißen.

ergibt sich aus (6) die allgemein-relativistische Formel zur Berechnung der Einweg-Geschwindigkeit

$$c_{\pm} = \frac{dl}{d\tau_{\pm}} = c + \frac{g_{0\alpha} dx^{\alpha}}{d\tau \sqrt{g_{00}}} = c \pm \frac{g_{0\alpha} |dx^{\alpha}|}{d\tau \sqrt{g_{00}}}. \tag{9}$$

Wie im Spezialfall der rotierenden Scheibe haben wir $c_{\pm}$ hier analog zur ursprünglichen Ableitung (4) als Quotient aus lokaler Eigenlänge und Eigenzeit bestimmt. Wir finden demzufolge

$$\Delta c = \pm \frac{g_{0\alpha} |dx^{\alpha}|}{d\tau \sqrt{g_{00}}}. \tag{10}$$

Man sieht sofort, daß die Einweg-Geschwindigkeit $c_{\pm}$ in allen statischen Gravitationsfeldern gleich $c$ ist, wie es auch sein muß. Ganz anders aber verhält sich die Angelegenheit, wenn es nicht mehr um statische, sondern um *stationäre* Systeme geht.

Gehen wir, um zu dem konkreten Beispiel der rotierenden Scheibe zurückzukehren, von den Zylinderkoordinaten $t, r, \varphi, z$ des Inertialsystems S über zu den entsprechenden Koordinaten des rotierenden Systems S* vermittels folgender Transformation, die interessanterweise genau einer GALILEI-Transformation, und im Ergebnis gerade den Bedingungen und Forderungen 1.) – 4.) und *) entspricht:

$$t = t^*, \ r = r^*, \ \varphi = \varphi^* + \omega t^*, \ z = z^*. \tag{11}$$

Damit erhalten wir

$$ds^2 = \left(1 - \frac{\omega^2 r^{*2}}{c^2}\right) c^2 dt^{*2} - 2\omega r^{*2} dt^* d\varphi^* - dr^{*2} - r^{*2} d\varphi^{*2} - dz^{*2}. \tag{12}$$

Für die mit natürlichen Uhren gemessenen Laufzeiten von Lichtsignalen folgt bei *angepaßter* Synchronisation aus (6) und (8)

$$d\tau_{\pm} = dt^*_{\pm} \sqrt{g_{00}} = \frac{dl}{c} \pm \frac{\omega r^{*2} |d\varphi^*|}{c^2 \sqrt{1 - \frac{\omega^2 r^{*2}}{c^2}}}. \tag{13}$$

Das räumliche Linienelement $dl$ ergibt sich hier bekanntlich zu

$$dl = \sqrt{dr^{*2} + \frac{r^{*2} d\varphi^{*2}}{1 - \frac{\omega^2 r^{*2}}{c^2}} + dz^{*2}}. \tag{14}$$

Schließlich aber finden wir aus der allgemeinen Beziehung (9) die Einweg-Geschwindigkeit $c_{\pm}$

$$c_{\pm} = \frac{dl}{d\tau_{\pm}} = c \pm \frac{g_{0\alpha} |dx^{\alpha}|}{d\tau \sqrt{g_{00}}} = c \pm \frac{\omega r^{*2}}{1 - \frac{\omega^2 r^{*2}}{c^2}} \frac{|d\varphi^*|}{c dt^*}. \tag{15}$$

Man liest sofort ab, daß die Einweg-Geschwindigkeit $c_{\pm}$ sowohl in radialer Richtung als auch in Richtung der $z^*$-Achse[17] wegen $d\varphi^* = 0$ einfach gleich $c$ ist, was natürlich zu erwarten war:

$$c_{\pm \, radial} = c_{\pm z^*} = c. \tag{16}$$

Zur Ermittlung der tangentialen Einweg-Geschwindigkeit $c_{\pm \, tangential}$ berechnen wir aus $ds^2 = 0$ zunächst einmal den *Koordinatenwert* $d\varphi^*/dt^*$

---

[17] Der Weg eines Lichtsignals kann auf einer rotierenden Scheibe natürlich nur auf *infinitesimalen* Strecken bzw. bei Verwendung hinreichend vieler Spiegel annähernd radial bzw. axial verlaufen.

$$\frac{d\varphi^*}{dt^*} = \frac{c}{r^*}\sqrt{1-\frac{dr^{*2}}{c^2 dt^{*2}}-\frac{dz^{*2}}{c^2 dt^{*2}}} \pm \omega \ . \tag{17}$$

Einsetzen in (15) ergibt dann unter der Voraussetzung $d\varphi^*/dt^* \neq 0$:

$$c_\pm = c \pm \frac{\omega r^* \left(\sqrt{1-\frac{dr^{*2}}{c^2 dt^{*2}}-\frac{dz^{*2}}{c^2 dt^{*2}}} \pm \frac{\omega r^*}{c}\right)}{1-\frac{\omega^2 r^{*2}}{c^2}} \ . \tag{18}$$

Und mit $dr^* = dz^* = 0$ findet man endlich:

$$c_{\pm \text{ tangential}} = \frac{c}{1 \mp \frac{\omega r^*}{c}} \ . \tag{19}$$

Dies stimmt wegen $v = \omega r$ und $r = r^*$ mit dem ursprünglich ohne Verwendung des mathematischen Apparats der allgemeinen Relativitätstheorie gefundenen Ausdruck (4) genau überein.

Welche der am Anfang des Abschnitts genannten verschiedenen Werte der Lichtgeschwindigkeit (s. Tab. 1) sind nun tatsächlich meßbar? Es ist sofort klar, daß sich nach Durchführung einer globalen Synchronisation zusätzlich zur lokalen Lichtgeschwindigkeit $c$ nun auch die längenbezogene Einweg-Lichtgeschwindigkeit $c_{l\pm}$ messen läßt, aus der sich dann die Einweg-Lichtgeschwindigkeit $c_\pm$ durch Berücksichtigung der zuvor erfolgten technischen Korrekturen der Ganggeschwindigkeit der natürlichen Uhren eindeutig ergibt. Da man nach EIN-STEIN „*darauf verzichten muß, den Koordinaten eine unmittelbare metrische Bedeutung zu geben (Koordinatendifferenzen = meßbare Längen bzw. Zeiten)*" [8], so kann in *Teilsystemen* auch der Koordinatenwert der Lichtgeschwindigkeit $c^*_\pm$ selbst nur von mittelbarer metrischer Bedeutung sein. Immerhin aber sind Unterschiede dieses Wertes – wie SHAPIRO [9] gezeigt hat – als Laufzeitverzögerungen $\Delta t^*$ *indirekt*[18]) meßbar.

## 9. GALILEI-Transformation und Relativitätstheorie

Die Transformation (11) vom ruhenden auf ein rotierendes System entspricht nicht etwa einer LORENTZ-, sondern exakt einer GALILEI-Transformation. *Da aber in der allgemeinen Relativitätstheorie jede Koordinaten-Transformation erlaubt ist, so muß natürlich auch eine GALILEI-Transformation erlaubt sein.* Es ist nun allerdings sehr bemerkenswert, daß ein entsprechender stationärer Übergang auf Basis einer LORENTZ-Transformation *nicht* möglich ist.

Für die eigentliche GALILEI-Transformation der Koordinaten $t = t^G$, $x = x^G + v t^G$, $y = y^G$, $z = z^G$ des Ruhsystems $S$ auf das bewegte Inertialsystem $S^G$ lautet das Linienelement:

$$ds^2 = \left(1-\beta^2\right)c^2 dt^{G2} - 2\beta c\, dt^G dx^G - \left(dx^{G2} + dy^{G2} + dz^{G2}\right) . \tag{20}$$

Die räumliche Entfernung, so wie sie sich bei der Ermittlung mit natürlichen Einheitsmaßstäben in $S^G$ ergibt, finden wir gemäß (7) zu:

$$dl = \sqrt{\frac{dx^{G2}}{1-\beta^2} + dy^{G2} + dz^{G2}} \ . \tag{21}$$

Aus (8) ergibt sich für die Eigenzeit $\tau$ einer in $S^G$ ruhenden Uhr:

$$d\tau = dt^G \sqrt{1-\beta^2} \ . \tag{22}$$

---

[18]) Daß hier – allerdings mit global falsch gehenden natürlichen Uhren – unzweifelhaft Unterschiede der *absoluten* Systemzeit $t^*$ gemessen werden, könnte möglicherweise ein Grund dafür sein, daß EINSTEIN die Berechnung dieses Effekts gewissermaßen 'vergessen' hat.

Und schließlich beträgt die Lichtgeschwindigkeit für eine Ausbreitung parallel zur $x^G$-Achse im bewegten System wieder

$$c^G_\pm = \frac{c}{1 \mp \beta}, \qquad (23)$$

was sich aus (15) ergibt, wenn man gemäß $ds = 0$ berücksichtigt, daß $dx^G/dt^G = -v \pm c$ sein muß. Nach diesem Muster kann man also auch in der Relativitätstheorie mit der GALILEI-Transformation rechnen, was für *lokale* Inertialsysteme bei zweckmäßiger Wahl der Systemkoordinaten in der Regel sogar unvermeidlich ist.

---

*Koordinatenwert* $c^*_\pm = dl^*/dt^*_\pm$ der Lichtgeschwindigkeit
    die auf die *Systemkoordinaten* bezogene Lichtgeschwindigkeit bei *globaler* Synchronisation. Auf der rotierenden Scheibe:

$$c^*_{\pm\,\text{tangential}} = \frac{r^* d\varphi^*}{dt^*} = c \pm \omega r^*$$

$$c^*_{\pm\,\text{radial}} = \frac{dr^*}{dt^*} = c\sqrt{1 - \frac{\omega^2 r^{*2}}{c^2}}.$$

*lokale* Lichtgeschwindigkeit $c = dl/d\tau$
    die mit *natürlichen* Maßstäben *und* Uhren gemessene Lichtgeschwindigkeit bei *lokaler* EINSTEIN-Synchronisation

$$\text{Naturkonstante} \quad c = \frac{1}{\sqrt{\varepsilon_o \mu_o}}$$

Einweg-Geschwindigkeit $c_\pm = dl/d\tau_\pm$
    die mit *natürlichen* Maßstäben *und* Uhren gemessene Lichtgeschwindigkeit bei *lokal angepaßter* Synchronisation. Entspricht dem auf ‚natürliche' Maßeinheiten umgerechneten Koordinatenwert der Lichtgeschwindigkeit. Auf der rotierenden Scheibe:

$$c_{\pm\,\text{tangential}} = \frac{dl}{d\tau_\pm} = \frac{c}{1 \mp \frac{\omega r^*}{c}}.$$

$$c_{\pm\,\text{radial}} = \frac{dl}{d\tau_\pm} = c$$

*längenbezogene* Einweg-Geschwindigkeit $c_{l\pm} = dl/dt^*_\pm$
    der auf *natürliche Maßstäbe,* aber *technische Systemuhren* bezogene Wert der Lichtgeschwindigkeit. In unserem Beispiel:

$$c_{l\pm\,\text{tangential}} = \frac{dl}{dt^*} = \frac{r^* d\varphi^*}{dt^* \sqrt{1 - \frac{\omega^2 r^{*2}}{c^2}}} = c\sqrt{\frac{1 \pm \frac{\omega r^*}{c}}{1 \mp \frac{\omega r^*}{c}}}.$$

$$c_{l\pm\,\text{radial}} = \frac{dl}{dt^*} = \frac{dr^*}{dt^*} = c\sqrt{1 - \frac{\omega^2 r^{*2}}{c^2}}$$

*zeitbezogene* Einweg-Geschwindigkeit $c_{\tau\pm} = dl^*/d\tau_\pm$
    der auf *natürliche Uhren,* aber *räumliche Systemkoordinaten* bezogene Wert der Lichtgeschwindigkeit bei *lokal angepaßter* Synchronisation. Auf der rotierenden Scheibe:

$$c_{\tau\pm\,\text{tangential}} = \frac{dl^*}{d\tau_\pm} = \frac{r^* d\varphi^*}{d\tau_\pm} = c\sqrt{\frac{1 \pm \frac{\omega r^*}{c}}{1 \mp \frac{\omega r^*}{c}}}.$$

$$c_{\tau\pm\,\text{radial}} = \frac{dl^*}{d\tau_\pm} = \frac{dr^*}{d\tau_\pm} = c$$

Tab. 1: Die verschiedenen Werte der Lichtgeschwindigkeit in der Übersicht

## 10. Das kosmische Bezugssystem: der wahre Raum und die wahre Zeit

Müssen natürliche Uhren notwendigerweise die *wahre* Zeit (natürliche Maßstäbe den *wahren* Raum) anzeigen, wenn sie, an ein und demselben Ort unter denselben Bedingungen nebeneinander ruhend, immer und überall die gleiche Ganggeschwindigkeit (bzw. die gleiche Länge) aufweisen? – Nein.

„ ... Ebensowenig kann man in K' eine den physikalischen Bedürfnissen entsprechende Zeit einführen, welche durch relativ zu K' ruhende, gleich beschaffene Uhren angezeigt wird." – *[10], S. 775* (die Bezeichnung *K'* meint ein gleichförmig rotierendes Koordinatensystem)

Mit der hier aus dem EHRENFEST'schen Paradoxon der rotierenden Scheibe von EINSTEIN – im Anschluß[19]) an KALUZAS Behandlung – gefolgerten Notwendigkeit, in die allgemeine Relativitätstheorie eine Systemzeit $t^*$ einzuführen, die nicht übereinstimmt mit der von *natürlichen* Uhren angezeigten Eigenzeit $\tau$, und Systemkoordinaten $x^{*\alpha}$, die sich in ihrer Gesamtheit nicht darstellen lassen durch entsprechende mit *natürlichen* Maßstäben angezeigte Längen $l^{\alpha}$, sind nach unserem Verständnis nur wenige Jahre nach Formulierung der speziellen Relativitätstheorie Raum und Zeit als *absolute* Größen in die Physik zurückgekehrt.

Diese Auffassung wird von vorneherein durch folgende Tatsache nahegelegt: Ganz unabhängig von der jeweils aktuellen Theorie über Zustand bzw. Entwicklung des Universums – mit Hilfe des DOPPLER-Effekts läßt sich statistisch immer ein ausgezeichnetes *Ruhsystem* festlegen, und zwar durch die Forderung größtmöglicher Isotropie des kosmischen Hintergrunds[20]). Die absoluten Geschwindigkeiten von Sonne und Erde sind auf dieser Basis bekanntlich längst ermittelt.

Die *spezielle* Relativitätstheorie zeigt nun zwar, daß in Inertialsystemen trotz Längenkontraktion und Zeitdilatation eine solche Koordinatenwahl möglich ist, bei der den Differenzen räumlicher Koordinaten unmittelbar meßbare Entfernungen, und den Differenzen der Zeitkoordinaten unmittelbar meßbare Zeiten entsprechen. Doch die *allgemeine* Relativitätstheorie zeigt gerade, daß dies in Nicht-Inertialsystemen bzw. bei Berücksichtigung der Gravitation nicht mehr möglich ist. Angesichts der Tatsache, daß es abgesehen vom kosmischen Ruhsystem überhaupt nur lokale Inertialsysteme geben kann, beweist dies also die Existenz eines *absoluten* Raums und einer *absoluten* Zeit – durchaus im Sinne NEWTONs – als notwendige Voraussetzung für die Beschreibung der physikalischen Wirklichkeit[21]).

Daß die wahren Systemkoordinaten allerdings in *lokalen* Teilsystemen nicht eindeutig identifizierbar sind, kann ihre Existenz ebensowenig widerlegen, wie die Eigenschaft einer Ebene, flach zu sein, durch die notwendige Bezugnahme auf weitgehend willkürlich wählbare krummlinige Koordinaten bei Verwendung temperaturabhängiger Maßstäbe widerlegt werden kann.

Den hier[22]) gezogenen Schlüssen über die wahre Bedeutung der Systemkoordinaten entsprechend, lassen sich alle herkömmlichen Aussagen über die ‚Raum-Zeit' der Relativitätstheorie unseres Erachtens am einfachsten verstehen als Aussagen über reale Objekte, Felder, Maßstäbe und Uhren, die im absoluten euklidischen Raum und in der absoluten kosmischen Zeit dem Einfluß von *Gravitationspotential* und *Bewegung* unterliegen. Im Hinblick auf die kosmischen Systemkoordinaten aber gilt:

*Raum und Zeit selbst sind keine physikalischen Objekte, denen sich veränderliche Eigenschaften zuschreiben lassen. Gegenstand der physikalischen Beschreibung sind allein Veränderungen gegenüber dem, was notwendigerweise unveränderlich ist, und dessen Unveränderlichkeit keiner Erklärung bedarf.*

---

[19]) s. Fußnoten [32/33])

[20]) Dies kann heute durch die Forderung größtmöglicher Isotropie der Hintergrundstrahlung definiert werden. Prinzipiell hätte man sich bereits mit HUBBLEs Entdeckung auf eine maximale Isotropie der beobachtbaren statistischen Verteilung der Rotverschiebung beziehen können, vorher auf mittlere Sterngeschwindigkeit Null. Und selbst wenn sich der uns heute bekannte Kosmos eines Tages als Teil eines Universums aus ähnlichen und anderen Gebilden zeigen sollte – es läßt sich immer ein ausgezeichnetes Ruhsystem finden (ein ernsthaftes Problem entstünde umgekehrt erst dann, falls es mehr als ein einziges derartiges Ruhsystem geben sollte).

[21]) Diese Auffassung von Raum und Zeit führt unmittelbar zu der Möglichkeit eines *stationären* kosmischen Linienelements im Rahmen der allgemeinen Relativitätstheorie. Auf die diesbezügliche Arbeit, aus der sich auch ein natürlicher Erklärungsansatz für den Pioneer-Effekt ergibt, wurde bereits in Fußnote [10]) hingewiesen.

[22]) s. auch Abschnitt f) des Anhangs

Was nun ist die wahre Zeit? Wenn die Systemzeit eines beliebigen abgeschlossenen Teilsystems dessen interne globale Zeit darstellt, dann ist die singularitätsfreie Systemzeit des Universums $t^*$ als die wahre *kosmische* Zeit zu verstehen. Es läßt sich zeigen[23], daß eine besonders einfache Synchronisation entsprechender im absoluten kosmischen Bezugssystem ruhender technischer Systemuhren im Sinne des Abschnitts 6 auf Basis der Reflexion im Zeitmittelpunkt prinzipiell immer möglich ist. *Die kosmische Zeit ist durch die Bedingung einer – abgesehen von kleinen Abweichungen in der Nähe lokaler Inhomogenitäten des Gravitationsfeldes – konstanten kosmischen Lichtgeschwindigkeit c eindeutig festgelegt.*

## 11. Neue Interpretation alter Versuche

Die Orts- und Richtungsabhängigkeit der Einweg-Lichtgeschwindigkeit in rotierenden Systemen ist experimentell längst bestätigt. Das wird sofort klar, wenn man darauf besteht, die Versuche von SAGNAC [11] sowie von MICHELSON und GALE [12] *systemintern*, d.h. ohne Rückgriff auf ein übergeordnetes Inertialsystem zu erklären. Diese Interferenzversuche haben bekanntlich folgende Näherungsformel für die Streifenverschiebung in rotierenden Systemen bestätigt:

$$\Delta Z \approx \frac{4\omega A}{c\lambda}. \tag{24}$$

Uns kommt es hier ganz allein darauf an, daß im rotierenden System $S^*$ bei gleichem Weg $U^*$ zweier peripher umlaufender Lichtstrahlen überhaupt keine Verschiebung der Interferenzstreifen auftreten dürfte, wäre die Geschwindigkeit der beiden Lichtsignale tatsächlich gleich groß. Da nämlich die mit einer einzigen Uhr gemessenen Frequenzen der beiden zu $L^*$ zurückkehrenden Lichtanteile in $S^*$ wegen der bei gleichförmiger Rotation stationären Lichtwege mit der Emissionsfrequenz $f^*$ übereinstimmen müssen, hätten gleiche Geschwindigkeiten $c_\pm$ auch gleiche Phasen zur Folge, was insgesamt zu keiner Streifenverschiebung Anlaß geben könnte. Benutzt man die unterschiedlichen Wellenlängen[24]

$$\lambda^*_\pm = \frac{c_\pm}{f^*}, \tag{25}$$

so ergibt sich der relativistisch exakte Wert der Streifenverschiebung für unser anfängliches Beispiel unmittelbar als Differenz der Anzahl von Wellenlängen, die den beiden Umläufen entsprechen:

$$\Delta Z = \frac{U^*}{\lambda^*_-} - \frac{U^*}{\lambda^*_+}. \tag{26}$$

Mit (3), (4) ergibt sich aus (25), (26) nunmehr als endgültiger Wert für die Streifenverschiebung

$$\Delta Z = \frac{4\pi\, v r f^*}{c^2 \sqrt{1-\frac{v^2}{c^2}}}. \tag{27}$$

Wenn man zum Vergleich $A = \pi r^2$ und $\omega = v/r$ in die klassische Verschiebungsformel (24) einsetzt, erhält man einen Näherungswert, der sich auch durch direkte Abschätzung aus dem Laufzeitunterschied (1) ergibt

$$\Delta Z \approx \frac{c\Delta T}{\lambda} \approx \frac{4\pi v r}{c\lambda} = \frac{4\pi v r f}{c^2}. \tag{28}$$

Wie zu erwarten war, stimmt dieser Näherungswert mit dem exakten Wert (27) bis auf Größen dritter Ordnung in $v/c$ überein.

## 12. Das Experiment von HAFELE und KEATING

Bekanntlich haben HAFELE und KEATING [13], ausgestattet mit Atomuhren, Rundflüge um die Erde sowohl in Ost-West-Richtung als auch in West-Ost-Richtung unternommen und dabei folgende Formel bestätigt gefun-den, welche die Zeitunterschiede zwischen den Uhren im Flugzeug und der ortsfesten Uhr auf der Erde näherungsweise beschreibt:

$$\tau - \tau_0 \approx \frac{2\pi R}{c^2}\left[\frac{gh}{v} \pm R\Omega - \frac{v}{2}\right]. \tag{29}$$

---

[23] Der entsprechende Nachweis wird in der bereits erwähnten Arbeit d. Verf. erbracht (s. Fußnote [10]).

[24] Dieser Unterschied in der Wellenlänge bedeutet nicht etwa, daß es verschiedenfarbiges Licht gleicher Frequenz geben sollte. Er verschwindet, wenn man von der globalen Synchronisation zur lokalen EINSTEIN-Synchronisation übergeht.

Diese Formel ergibt sich sehr einfach durch Rechnung im rotationsfreien Schwerpunktsystem, d.h. bei *externer* Betrachtung des rotierenden Systems Erde. Dabei ist $R$ der Radius des überflogenen Breitengrads, $\Omega$ die Winkelgeschwindigkeit der Erde, $h$ die Flughöhe über dem Meeresspiegel und $v$ die Fluggeschwindigkeit relativ zur Erdoberfläche. Die mittleren Meßwerte betrugen -59 ns für West-Ost-Flug und 273 ns für Ost-West-Flug.

Betrachten wir nun den theoretischen Grenzfall $h = 0$ und $v = 0$, auf den die Autoren nicht eingegangen sind. In der Praxis wäre das etwa ein langsamer Transport per Schiff. Der resultierende Zeitunterschied wäre bei Rückkehr der rund um die Erde transportierten Uhr nicht etwa Null. Vielmehr ergibt sich aus (29) ein Zeitunterschied von $2\pi R^2\Omega/c^2$, was gerade KALUZAS ‚Schlußfehler' entspricht. Und genau hier gibt es einen Hund, der nicht bellte: Wäre nämlich jemand auf die Idee verfallen, den Gang der Uhren im Flugzeug während der Erdumrundung mit der zurückgelassenen, ortsfesten Uhr durch Signalaustausch auf Grundlage des EINSTEIN'schen Prinzips der Reflexion im Zeitmittelpunkt zu vergleichen, wie dies in ALLEYs Maryland-Experiment [6] tatsächlich geschehen ist, so hätte er – von technischen Schwierigkeiten abgesehen – gravierende Abweichungen[25] in der Größenordnung von ±160 ns zwischen den auf diese Weise gefundenen Meßwerten und den theoretischen Werten feststellen müssen. Die so gemessenen Zeitunterschiede wären für Ostflug und Westflug die gleichen gewesen, beide nämlich in der Größenordnung von +100 ns, was sich aus (29) ergibt, wenn man jeweils das mittlere Glied der Klammer wegläßt.

Daß aber die tatsächlichen Meßwerte von HAFELE und KEATING eben einmal -59 ns, das andere Mal +273 ns betrugen, widerlegt die Tauglichkeit des EINSTEIN'schen Prinzips der Reflexion im Zeitmittelpunkt für die globale Synchronisation – und damit die Konstanz der Einweg-Lichtgeschwindigkeit auf der Erde.

## 13. Zur Definition des Meters

Vom Comité International des Poids et Mesures wird als erstes von zwei möglichen Verfahren ausdrücklich die praktische Festlegung des Meters auf Basis des Zusammenhangs $l = c \cdot t$ empfohlen [1]. Dies impliziert aber die Aussage, daß die Länge des Äquators z.B. prinzipiell mit beliebiger Genauigkeit durch die Angabe mitgeteilt werden kann, welchen Bruchteil einer Sekunde elektromagnetische Wellen brauchen, diesen zu umlaufen. Gerade das aber ist ganz sicher unmöglich[26]. Denn welche Korrekturen auch immer man anbringen mag, es bleibt ein richtungsabhängiger Unterschied der Lichtgeschwindigkeit von bis zu ca. ±460 m/s (lokale Rotationsgeschwindigkeit) an ein und demselben Ort der Erde, wenn man zur Abmessung der Laufzeit Uhren verwendet, die *global* richtig synchronisiert sind. Die Beibehaltung der aktuellen Festlegung des Meters hätte also insbesondere die unsinnige Konsequenz, daß sich der in östlicher Richtung gemessene Erdumfang am Äquator von dem in westlicher Richtung gemessenen um ca. 2*60 m unterscheiden müßte. Zu sagen, die Einweg-Lichtgeschwindigkeit auf der rotierenden Erde sei konstant, hat tatsächlich keine größere Berechtigung als die Aussage, die Erde sei eine Scheibe. Denn praktisch meßbare Abweichungen treten hier wie dort erst in globalen Dimensionen auf.

Natürlich kann man Uhren zur Festlegung des Meters verwenden, die aufgrund einer vorgeschriebenen Synchronisation lokal das gewünschte Ergebnis liefern. Eine unmißverständliche Definition des Meters aber – wenn sie überhaupt auf der Geschwindigkeit des Lichts und nicht auf der Länge stehender Lichtwellen basieren soll – müßte nach den Ergebnissen der vorausgegangenen Abschnitte entweder ausdrücklich auf die EINSTEIN'sche (oder eine äquivalente) Synchronisationsvorschrift Bezug nehmen, oder aber sie könnte unabhängig von jeder Synchronisation lauten: *Das Meter ist die Länge der Strecke, die von Licht im Vakuum während der Zeitspanne von 2 mal 1/299792458 s hin und zurück durchlaufen wird.*

---

[25]) Bei ALLEYs zitiertem Maryland-Experiment dagegen hätten sich entsprechende Abweichungen nicht bemerkbar machen können, weil hier nur eine relativ kleine Schleife mit geringer Ost-West-Komponente geflogen wurde.

[26]) Etwas anderes ist es, Entfernungen in Lichtsekunden $cs$ anzugeben, wenn man diese dabei als zusammengesetzt aus hin- und zurück durchlaufenen *Teilstrecken*, nicht aber auf Basis der Einweg-Lichtgeschwindigkeit versteht.

Die hier vorgeschlagene Formulierung geht nicht nur über den auf statische – und damit auch rotationsfreie – Systeme begrenzten Einsatzbereich der derzeitigen Definition hinaus, sie läßt sich in der Praxis überdies leicht umsetzen, indem man sich nach bewährtem Muster auf die Länge stehender Lichtwellen einer Referenzstrahlung gegebener oder gemessener Frequenz bezieht. Denn es ist zu beachten, daß sich die Abweichungen der Einweg-Lichtgeschwindigkeit vom Wert der Naturkonstanten $c$ in diesen Experimenten zur praktischen Festlegung eines Längenstandards nicht bemerkbar machen, weil hierbei die Interferenz ebener, auf demselben Weg hin- und zurücklaufender Wellen benutzt wird[27]. Insbesondere ist auch die Anzahl der Wellenknoten[28] um den Äquator (oder auf der Peripherie einer rotierenden Scheibe) grundsätzlich unabhängig von der Frage der lokalen oder globalen Synchronisation – und damit vom Wert der Einweg-Lichtgeschwindigkeit schlechthin. Es ist zwar richtig,

„ ... that these various forms, making reference either to the path travelled by light in a specified time interval or to the wavelength of a radiation of measured or specified frequency, have been the object of consultations and deep discussions, have been recognized as being equivalent ... " – *[1]*.

Doch die naturgemäß stationäre *Länge stehender Wellen* als doppelter Abstand zweier Schwingungsknoten ist nicht das gleiche wie die Wellenlänge als Abstand zweier *gleichzeitig* lokalisierter benachbarter Wellenberge, da letztere genau wie die Einweg-Lichtgeschwindigkeit abhängig ist vom gewählten Synchronisationsverfahren gemäß $\lambda_\pm = c_\pm/f$, erstere aber nicht.

### Anhang: Das EHRENFEST'sche Paradoxon und die Unmöglichkeit einer scharfen Trennung zwischen relativistischer Kinematik und Dynamik

Die NEWTON'sche Kinematik des *starren Körpers* ist für die klassische Physik bekanntlich von fundamentaler Bedeutung. Nur mit ihrer Hilfe ist es möglich, die Bewegungsgesetze ausgedehnter Körper unter dem Einfluß äußerer Kräfte auf eine Mechanik des Massenpunkts zu reduzieren. Wenn etwas entsprechendes auch in der Relativitätstheorie möglich sein soll, ist es jedenfalls ganz unverzichtbar, das Modell des starren Körpers, wenn auch in abgewandelter Form, als *stationären Körper* hierhinein zu übertragen. Dabei läßt es sich aber keineswegs von Anfang an ausschließen, daß die dazu notwendigen Einschränkungen des Starrheitsbegriffs auch Konsequenzen für den Begriff des Massenpunkts haben. Denn natürlich gibt es keine punktförmigen Teilchen, jeder Körper ist ausgedehnt.

Was könnte nun – angesichts andernfalls auftretender Überlichtgeschwindigkeiten – an die Stelle des Begriffs ‚starrer Körper' in einer relativistischen Mechanik treten? Die oben immer wieder benutzte Voraussetzung lokaler Inertialsysteme wurde seinerzeit von BORN [14] zur Grundlage der Definition eines *relativ-starren*[29] Körpers gemacht. Wegen der Längenkontraktion kann dies – in einer äquivalenten Formulierung von EHRENFEST – nur bedeuten, daß für einen solchen Körper

„ ... jedes seiner infinitesimalen Elemente in jedem Moment für einen ruhenden Beobachter gerade diejenige LORENTZ-Kontraktion (gegenüber dem Ruhezustand) aufweist, welche der Momentangeschwindigkeit des Elementmittelpunktes entspricht". – *[4]*

---

[27]) Die Ausbildung stehender Wellen wird hier keineswegs dadurch gestört, daß bezogen auf das stationäre System S* gemäß (25) Licht gleicher Frequenz $f$, aber verschiedener Wellenlänge $\lambda_\pm = c_\pm/f$ überlagert wird. Bei Berücksichtigung des Ausdrucks (4) für die Einweg-Geschwindigkeit $c_\pm$ (s. auch Tab. 1) ergibt sich nämlich mit Wellenzahl $k = 2\pi/\lambda$, Kreisfrequenz $\omega = 2\pi f$, Amplitude $A$ aus den Anteilen $a_\pm = A \sin(\omega t \pm k_\pm x)$ durch Überlagerung $a = 2A \cos kx \,[\cos(kxv/c)\sin\omega t + \sin(kxv/c)\cos\omega t]$, sodaß die Schwingungsknoten nach wie vor an der gleichen Stelle liegen wie bei Überlagerung von Licht derselben Frequenz $f$ und einheitlicher Wellenlänge $\lambda = c/f$ (das kann das auch gar nicht anders sein, wie sich aus den nächsten Sätzen oben ohne jede Rechnung von selbst ergibt).

[28]) Gelegentlich wird in diesem Zusammenhang übrigens die Frage gestellt, ob nun das $E$-Feld oder das $B$-Feld der richtige Repräsentant der Lichtwelle sei. Tatsächlich aber kann dies nur ein Knoten der ‚LORENTZ-Feldstärke' $E + [v/c \times B]$ sein (Quotient aus LORENTZ-Kraft und Probeladung). Im Unterschied zum $E$- oder $B$-Feld ist nämlich allein die räumliche Zuordnung von Knoten der ‚LORENTZ-Feldstärke' zu bestimmten Raumpunkten eines Inertialsystems lorentzinvariant.

[29]) in der Bezeichnung von EHRENFEST (BORNs eigene Bezeichnung ist ‚starr')

Ebenso wie der starre Körper der NEWTON'schen Mechanik soll also auch der relativ-starre Körper der Relativitätstheorie zwar nicht dynamisch, im Unterschied zur NEWTON'schen Mechanik aber *kinematisch* deformierbar sein. EHRENFEST fährt fort:

> „Als ich mir vor einiger Zeit die Konsequenzen dieses Ansatzes veranschaulichen wollte, stieß ich auf Folgerungen, die zu zeigen scheinen, daß o b i g e r  A n s a t z  s c h o n  f ü r  e i n i g e  s e h r  e i n f a c h e  B e w e g u n g s t y p e n  z u  W i d e r s p r ü c h e n  f ü h r t." *(Hervorhebung wie im Original)* − [4]

Er schließt dies daraus, daß der Radius $R_{rot}$ einer relativ-starren rotierenden Scheibe einerseits, weil immer senkrecht zur Bewegungsrichtung, gleich groß sein müßte wie der Radius $R$ derselben Scheibe, bevor sie in Rotation versetzt wurde, andererseits aber zugleich auch kleiner, weil der Umfang $U_{rot} = 2\pi R_{rot}$ der rotierenden Scheibe aufgrund der FITZGERALD-LORENTZ-Kontraktion für einen im Inertialsystem $S$ ruhenden Beobachter verkürzt sei. EHRENFEST hat damit schlüssig bewiesen, daß es einen relativ-starren *makroskopischen* Körper nicht geben kann. Und das ist der Beginn einer langen Auseinandersetzung mit dem nach seinem Entdecker benannten Paradoxon, wobei die rotierende Scheibe für die Relativitätstheorie unseres Erachtens eine ähnliche Rolle spielt wie der schwarze Körper für die Quantenmechanik.

**a.    BORNs Interpretation**

In der Fußnote einer späteren Arbeit von 1910 interpretiert BORN denselben Sachverhalt dahingehend, EHRENFEST habe gezeigt,

> „ ... daß ein ruhender Körper niemals in gleichförmige Rotation gebracht werden kann; dieselbe Tatsache hatte ich schon mit Herrn A .  E i n s t e i n auf der Naturforscherversammlung in Salzburg besprochen:" *(Hervorhebung wie im Original).* − [15]

Gemeint ist hier ein relativ-starrer Körper. Nach dem Hinweis, daß ein solcher dementsprechend „*nicht die erforderlichen 6 Freiheitsgrade aufzuweisen hat*", sondern nur drei, weil die drei Freiheitsgrade der Rotation entfallen, verteidigt BORN die Verwendbarkeit seines Starrheitsbegriffs „*zur Grundlegung der Dynamik der Elektronen*" mit dem damals noch richtigen Hinweis: „*Es gibt überhaupt keine Erscheinung, zu deren Erklärung bislang Rotationen der Elektronen herangezogen worden sind.*" Die Erfahrungstatsache, daß gewöhnliche materielle Körper sehr wohl in Rotation versetzt werden können, erklärt er so:

> „Denkt man sich die materiellen Körper aus Atomen und Elektronen aufgebaut, die etwa in meinem Sinne ‚starr' sind, so werden diese, wenn der ganze Körper rotiert, krummlinige Bahnen beschreiben, die mit ihrer Kinematik wohl verträglich sind ..." − [15]

Wenn aber das Verhalten rotierender makroskopischer Körper nur unter dieser Voraussetzung mit einer relativistischen Kinematik verträglich ist, dann läßt sich – BORNs Argumentation auf die Spitze getrieben – daraus ein relativistischer ‚Beweis' für die Existenz von Atomen ableiten: *Weil jeder beliebige makroskopische Körper in Rotation versetzt werden kann, was wegen des EHRENFEST'schen Paradoxons nicht möglich wäre, wenn es sich bei diesem um ein relativ-starres Kontinuum handelte, folgt notwendig, daß ein makroskopischer Körper aus Atomen besteht.* Dies allerdings mit der hier vorausgesetzten Einschränkung, daß es sich bei einem makroskopischen Körper im Rahmen einer relativistischen Kinematik nicht um ein *elastisches* Kontinuum handeln kann.

Mehr noch: die Tatsache, daß ein zunächst ruhender relativ-starrer Körper niemals in gleichförmige Rotation gebracht werden kann, bedeutet umgekehrt, daß ein einmal rotierender relativ-starrer Körper niemals abgebremst werden kann. Wenn wir dementsprechend BORNs Argumentation aus heutiger Sicht gewissermaßen vom Kopf auf die Füße stellen, so könnte sich aus der zusätzlichen Hypothese, daß der Begriff ‚relativ-starr' nur auf mikroskopische, nicht aber auf makroskopische Körper anwendbar sei, ein Hinweis auf konstant bleibende Eigendrehimpulse, d.h. den *Spin* von Mikroteilchen ableiten lassen.

**b.    PLANCKs Unterscheidung**

Von PLANCK wurde das EHRENFEST'sche Paradoxon aufgegriffen, indem er die wichtige Unterscheidung macht zwischen der richtigen Aussage, daß ein Körper in Bewegung um den FITZGERALD-LORENTZ-Faktor ver-

kürzt[30]) erscheine, und der – in der Regel unzutreffenden – Aussage, daß ein Körper bei der Beschleunigung auf eine entsprechende Geschwindigkeit gerade eine Verkürzung um den FITZGERALD-LORENTZ-Faktor erfahre[31]). Zum Problem des starren Körpers bemerkt er dann:

> „ ... scheint mir der Versuch, die für die gewöhnliche Mechanik so wichtige Abstraktion des starren Körpers auch für die Relativitätstheorie fruchtbar zu machen, keinen rechten Erfolg zu versprechen." − *[16]*

In der Beschleunigung eines Körpers etwa vom Zustand der Ruhe auf gleichförmige Rotation und der daraus resultierenden Deformation sah er insbesondere nicht eigentlich ein kinematisches, sondern vor allem ein ‚elastisches' Problem.

### c. V. LAUEs Beweis der Unmöglichkeit starrer Körper

Daß es überhaupt weder starre noch relativ-starre makroskopische Körper geben kann, wurde 1910 durch V. LAUE [17] in voller Allgemeinheit bewiesen. Ein makroskopischer Körper nämlich, der gleichzeitig an beliebig vielen Stellen angestoßen werden kann, muß – im Unterschied zum starren bzw. relativ-starren Körper – beliebig viele Freiheitsgrade haben, weil sich alle Störungen höchstens mit Lichtgeschwindigkeit ausbreiten, und sich demzufolge für eine gewisse, wenn auch sehr kurze, Zeitspanne nach dem Anstoß nicht gegenseitig beeinflussen können.

### d. KALUZAs Einführung der nichteuklidischen Geometrie

Bereits 1910 hat KALUZA [5] – in einer in Bezug auf die mathematische Behandlung bahnbrechenden Arbeit – nicht nur auf „die theoretische Möglichkeit eines Nachweises der Erdrotation durch rein optische bzw. elektromagnetische Experimente" hingewiesen, wobei er einen ‚Schlußfehler' der Synchronisation akzeptiert und diesen mit maximal $2*10^{-7}$ s beziffert (s Abschn. 5), sondern vor allem auch – drei Jahre vor EINSTEIN und GROSSMANN [18] – die *nichteuklidische Geometrie* in die Relativitätstheorie eingeführt[32]). Und zwar fließen diese Konsequenzen daraus, daß er die Gültigkeit der speziellen Relativitätstheorie im übergeordneten Inertialsystem voraussetzt und auf die lokalen Inertialsysteme des rotierenden Systems anwendet, ohne auf den Übergang von der zuvor ruhenden zur gleichförmig rotierenden Scheibe einzugehen. Das Verhältnis von Kreisumfang $U^*$ zu Radius $R^*$ einer rotierenden Scheibe $S^*$ ergibt sich demzufolge für einen mitbewegten Beobachter *größer* als $2\pi$, was nach KALUZA ausdrücklich keinen Widerspruch zur Relativitätstheorie bedeutet.

### e. EINSTEINs Übertragung auf das Gravitationsfeld

KALUZAs richtige Antwort $U^*/R^* > 2\pi$ ist nun aber ganz unabhängig davon, welche Deformation die Scheibe $S^*$ tatsächlich dadurch erfahren hat, daß sie allmählich in Rotation versetzt wurde. Inhaltlich sind es gerade KALUZAs Ergebnisse[33]), die EINSTEIN später sagen lassen:

> „In der allgemeinen Relativitätstheorie können Raum- und Zeitgrößen nicht so definiert werden, daß räumliche Koordinatendifferenzen unmittelbar mit dem Einheitsmaßstab, zeitliche mit einer Normaluhr gemessen werden könnten." − *[10], S. 775*

Das EHRENFEST'sche Paradoxon beweist hier nach unserem Verständnis die Ungültigkeit einer *mit deformierten, natürlichen Maßstäben betriebenen* euklidischen Geometrie zunächst im rotierenden Bezugssystem – dann aber mit EINSTEINs genialem *Äquivalenzprinzip* auch für jedes Gravitationsfeld überhaupt. Ist das EHRENFEST'sche Problem der rotierenden Scheibe damit vollständig gelöst?

### f. Die rotierende Scheibe im euklidischen Raum

Das Problem der rotierenden Scheibe hat EINSTEIN schließlich zu der Auffassung geführt, daß der dreidimensionale Raum durch das wahre Gravitationsfeld ‚gekrümmt'[34]) sei. Gerade dieses Problem zeigt aber umge-kehrt sehr einfach, daß sich aus der Tatsache eines nichtverschwindenden räumlichen Krümmungstensors – wie ihn KALUZAs Linienelement (14) impliziert – keineswegs eine ‚Krümmung des dreidimensionalen Raums' ableiten läßt. Denken wir uns nämlich zwei in einem gewissen Abstand voneinander mit unterschiedlicher Winkelgeschwindigkeit rotierende Scheiben, deren Drehachsen beide im Vakuum desselben Inertialsystems ruhen. Welche ‚Krümmung' sollte sich für den dreidimensionalen Raum zwischen den Scheiben wohl ergeben − die aus dem räumlichen Linienelement der einen oder die aus dem der anderen Scheibe resultierende? Es ist klar, daß es auf diese Frage nur eine einzige sinnvolle Antwort gibt: beide Scheiben rotieren im absoluten Raum[35]) und dieser ist *euklidisch*.

---

[30]) Daß sich diese Verkürzung im Auge des Beobachters bekanntlich als Drehung darstellen müßte, tut hier nichts zur Sache.

[31]) In diesem Sinne hat sich später auch EINSTEIN geäußert.

[32]) Die gesamte Arbeit hat einen Umfang von einer Druckseite plus drei Halbzeilen (s. auch Abschn. 5).

[33]) Daß KALUZAs Arbeit von EINSTEIN unseres Wissens später nicht zitiert worden ist – ebensowenig wie die von EHRENFEST u.a. – läßt keineswegs darauf schließen, daß er sie nicht gekannt hat. Andererseits hat KALUZA das von ihm behandelte Problem der rotierenden Scheibe ganz sicher nicht im Zusammenhang mit der Gravitation gesehen.

Zwar könnte man einwenden, der Raum werde nur durch *wahre* Gravitationsfelder ($R^i_{klm} \neq 0$) gekrümmt. Doch auch diese Position erweist sich als unhaltbar, wenn man daran festhält, daß sich physikalische Abläufe prinzipiell vollständig und lückenlos in Raum und Zeit beschreiben lassen. In Wahrheit nämlich beweist gerade das Auftreten eines nichteuklidischen räumlichen Linienelements die Tatsache, daß insbesondere die Beschreibung von Vorgängen bei zeitlich veränderlichem Gravitationspotential nur im euklidischen Raum möglich ist. Dies ergibt sich daraus, daß es keine Koordinaten-Transformation gibt, die von einer einmal gegebenen Krümmung des Raumes zu einer anderen führen kann. Doch die lückenlose deterministische Beschreibung eines Bewegungsablaufs drückt sich mathematisch gerade durch eine entsprechende Koordinaten-Transformation aus! Und eine solche ist eben nur im euklidischen Raum möglich. Alle zeitlichen Veränderungen eines wahren makroskopischen *Kontinuums* wären dabei zwangsläufig verbunden mit dem spontanen Auftreten von Rissen und Sprüngen[36]).

*Umgekehrt folgt aus der Relativitätstheorie, daß man an einer Beschreibung realer Abläufe mit natürlichen Maßstäben und Uhren – ohne Bezug auf den absoluten Raum und die absolute Zeit – nur um den Preis eines Verzichts auf eine vollständige deterministische Beschreibung festhalten kann.*

In den räumlichen Koordinaten des raum-zeitlichen Linienelements (12) sehen wir also Zylinderkoordinaten des dreidimensionalen euklidischen Raums. Der mathematische Apparat der nichteuklidischen Geometrie jedoch, der sich im räumlichen Linienelement (14) KALUZAs manifestiert, kommt dadurch ins Spiel, daß er den notwendigen Zusammenhang herstellt zwischen den absoluten Entfernungen des euklidischen Raums und den natürlichen, wie sie mit FITZGERALD-LORENTZ-kontrahierten, an der Rotation teilnehmenden materiellen Meterstäben – bzw. mit ‚Lichtmetern' – gemessenen werden[37]).

Alle Aussagen über die ‚Raum-Zeit' der Relativitätstheorie lassen sich unseres Erachtens – etwa im Sinne des LORENTZ'schen *dynamischen* Ansatzes [19], der bekanntlich auch von POINCARÉ [20] vertreten wurde[38]) – am einfachsten verstehen als Aussagen über reale Maßstäbe und Uhren, die im absoluten euklidischen Raum und in der absoluten kosmischen Zeit dem Einfluß von Gravitationspotential und Bewegung unterliegen. Damit werden die beiden fundamentalen Postulate der speziellen Relativitätstheorie nun einfach zu heuristischen Prinzipien[39]):

a) *Natürliche Uhren lassen sich lokal derart synchronisieren, daß die Naturgesetze in allen Inertialsystemen bei Verwendung natürlicher Maßstäbe die gleiche Form annehmen.*

---

[34]) Sollte der Begriff der ‚Krümmung' in diesem Zusammenhang lediglich in geometrischer Analogie auf den *mathematischen* Sachverhalt eines nichtverschwindenden räumlichen ‚Krümmungs'-Tensors hinweisen, so wäre dies nichts weiter als ein – allerdings unnötig Verwirrung stiftendes – Sprachproblem. Doch EINSTEIN meinte insbesondere in seiner Kosmologie etwas anderes.

[35]) Man könnte noch versuchsweise einwenden, nur der von den Scheiben selbst eingenommene Raum sei gekrümmt, nicht aber der Zwischenraum. Doch ist dieser Einwand nicht stichhaltig, weil sich KALUZAs räumliches Linienelement aus dem Verhalten von Lichtstrahlen *außerhalb* der rotierenden Scheibe bestimmen läßt.

[36]) Und zwar spätestens dann, wenn wir – obwohl nur theoretisch – versuchen, die Härte des Materials im Grenzwert gegen Unendlich gehen zu lassen.

[37]) Nach den Ausführungen des Hauptteils dieser Arbeit ist klar, daß beide Meßmethoden auf das gleiche hinauskommen, wenn wir unter dem ‚Lichtmeter' einen Längenstandard verstehen, der auf der Konstanz der durchschnittlichen lokalen Lichtgeschwindigkeit für Hin- und Rückläufe beruht.

[38]) Der Einfluß POINCARÉs auf die Entwicklung des Zeitbegriffs gerade im Sinne der späteren EINSTEIN'schen speziellen Relativitätstheorie wird unseres Erachtens in heutigen Lehrbüchern nicht ausreichend gewürdigt.

[39]) Diese Formulierung der beiden fundamentalen Postulate der speziellen Relativitätstheorie betont natürlich eher die Auffassung von LORENTZ und POINCARÉ.

*b) Bei dieser speziellen Synchronisation ist die Einweg-Lichtgeschwindigkeit in einem beliebigen Inertialsystem gleich der Naturkonstanten c.*

Längen- und Zeitvergleich werden wieder transitiv, die FITZGERALD-LORENTZ-Kontraktion wird – wie es das Kausalitätsprinzip verlangt – von einer kinematischen Deformation wieder zu einer dynamischen, und die Energiedichte des Gravitationsfeldes schließlich wird lokalisierbar[40]).

**g. Längenkontraktion – Kinematik oder Dynamik?**

Fest steht ohne jeden Zweifel, daß es *makroskopische* Körper gibt, die sich in Rotation versetzen lassen. Denken wir uns ein 314-eckiges ‚Rad' aus lauter gleichen Maßstäben zusammengefügt, dessen 314 Speichen jede wiederum aus 50 zusammengefügten Maßstäben besteht. Der Quotient von Umfang und Durchmesser $U^*/d^*$ bleibt also für einen mitbewegten Beobachter in S* anscheinend immer gleich 314/100 (unsere Näherung für π) – ganz unabhängig, ob sich das Rad dreht oder nicht. Dies steht in eklatantem Widerspruch zu der Aussage: Die Maßstäbe, aus denen sich das ‚Rad' zusammensetzt, erleiden keine andere Deformation außer der FITZGERALD-LORENTZ-Kontraktion. Da nämlich KALUZAS Ergebnis $U^* > U$ sicher richtig ist, können sich die 314 Maßstäbe, aus denen der Umfang des ‚Rads' besteht, beim direkten Vergleich mit dem transportablen Einheitsmaßstab nicht als gleich lang erweisen wie dieser. Sie sind also nicht nur *kinematisch*, sondern zwangsläufig auch *dynamisch* deformiert. Daraus schließen wir: *Eine scharfe Trennung von relativistischer Kinematik und Dynamik ist prinzipiell unmöglich.* Daß es solch eine scharfe Trennung tatsächlich nicht geben kann, ist zwar längst bekannt – bisher allerdings nur aus der Quantenmechanik: *in HEISENBERGS Relation nämlich finden sich immer kinematische mit dynamischen Unschärfen zur PLANCK'schen Konstanten verknüpft.*

Wie nun verhalten sich makroskopische Körper, die allmählich in gleichförmige Rotation versetzt werden? Welche Auswirkungen hat es, wenn man von leicht deformierbaren zu immer schwerer deformierbaren Materialien übergeht? Offenbar muß es für diese – abhängig von der jeweils erreichbaren Rotationsgeschwindigkeit – *relativistische* Grenzwerte der minimalen Deformation geben. Damit wird aber von der Relativitätstheorie zwangsläufig die Frage nach Materialkonstanten gestellt.

Gegenüber dem oben zitierten dynamischen Ansatz besagt der EINSTEIN'sche kinematische Ansatz sinngemäß: Längenkontraktion, Zeitdilatation und andere verwandte Phänomene ergeben sich unabhängig von der Natur irgendwelcher Kräfte allein aus der Kombination von Relativitätsprinzip mit dem Prinzip von der Konstanz der Lichtgeschwindigkeit und bedürfen keiner weiteren Erklärung.

Nach unserem Verständnis hätte EINSTEIN nicht recht mit diesem Verzicht auf eine dynamische Erklärung, wenn es richtig wäre, für die *gesamte* Physik an der Möglichkeit einer prinzipiell vollständigen, kausalen Beschreibung des Naturgeschehens festzuhalten[41]).

Aus der Tatsache nämlich, daß sich durchaus zutreffende Aussagen machen lassen über stationäre Körper in verschiedenen Bewegungs-Zuständen, ohne daß es aber möglich ist, vollständig zu beschreiben, wie sich solche Körper während entsprechender *Übergänge*, d.h. in Beschleunigungsphase verhalten, schließen wir, daß einer vollständigen, kausalen raumzeitlichen Beschreibung des Naturgeschehens bereits innerhalb der Relativitätstheorie prinzipielle Grenzen gesetzt sind. Bei konsequenter Berücksichtigung dieser Grenzen, die – wie oben gezeigt – die prinzipielle Unmöglichkeit einer scharfen Trennung von relativistischer Kinematik und Dynamik beinhalten, sollten sich schließlich die unterschiedlichen Ansätze von LORENTZ und POINCARÉ einerseits sowie EINSTEIN andererseits miteinander vereinbaren lassen:

Aus heutiger Sicht ist klar, daß die in diesem Anhang aufgeworfenen Fragen allesamt nur im Sinne der Quantenmechanik lösbar sind. So werden zusammengesetzte Körper zwischen verschiedenen, eng beieinanderliegenden Zuständen wechseln, wobei sich entsprechende Übergänge der lückenlosen raum-zeitlichen Beschreibung entziehen. Bereits mit der EINSTEIN'schen Zuordnung von Längenkontraktion und Zeitdilatation zur relati-vistischen *Kinematik* wurde unseres Erachtens der Verzicht auf eine vollständige, kausale Beschreibung des Naturgeschehens akzeptiert – und zwar lange vor der Quantenmechanik. In einer früheren Arbeit des Autors [22] wurde gezeigt, daß die relativistische Mechanik – im Gegensatz zu einer weit verbreiteten Auffassung – selbst in einfachsten Situationen nicht als bloß quantitativ modifizierte NEWTON'sche Mechanik verstanden werden kann. Dies ergibt sich vor allem daraus, daß die lokale Relativität der Gleichzeitigkeit wegen der Existenz von Umkehrpunkten in abgeschlossenen Systemen zwangsläufig auch *dynamische* Paradoxa[42]) mit sich bringt.

---

[40]) Den Übergang vom ausgezeichneten kosmischen Ruhsystem zu beliebigen Bezugssystemen leistet die bimetrische Formulierung der allgemeinen Relativitätstheorie durch ROSEN [21] u.a. auf Basis eines mathematischen Ansatzes von LEVI-CIVITA (s. a. Literaturangaben in [3], Bd. II §54).

[41]) Es ist sehr bemerkenswert, daß gerade EINSTEIN – als der spätere exponierte Verteidiger des Kausalitätsprinzips gegen die Kopenhagener Schule schlechthin – für FITZGERALD-LORENTZ-Kontraktion und Zeitdilatation nicht nur auf eine kausale Begründung verzichtet, sondern jeden Versuch einer kausalen Erklärung mehr oder weniger ablehnt.

**h. Abschließende Bemerkung zu diesem Anhang**

Das EHRENFEST'sche Paradoxon ist vom Ansatz her das Paradoxon der Längenkontraktion. Es soll deshalb zum Abschluß dieses Anhangs nicht unerwähnt bleiben, daß die Längenkontraktion im Unterschied zu beinahe allen anderen Effekten der speziellen Relativitätstheorie – insbesondere zu dem der Zeitdilatation – keineswegs in gleichem Maße als unmittelbar experimentell gesichert angesehen werden kann. Dies wird deutlich, wenn man es unternimmt, dieses Phänomen versuchsweise zu leugnen, indem man die Koordinate $x'$ des bewegten Systems S' in der üblichen LORENTZ-Transformation einfach gemäß $x' = X' / \sqrt{1-\beta^2}$ ersetzt. Dann zeigt sich, daß es möglich ist, an den allermeisten Resultaten der speziellen Relativitätstheorie zumindest näherungsweise festzuhalten, solange nur die Geschwindigkeiten der bewegten Inertialsysteme – nicht aber notwendigerweise die der darin enthaltenen Objekte – hinreichend klein sind. Zwar kann nun das Relativitätsprinzip nur noch näherungsweise gelten, d.h. es gibt in diesem Fall notwendigerweise ein ausgezeichnetes Bezugssystem, auch ist die Geschwindigkeit des von einer im bewegten System ruhenden Quelle ausgesandten Lichts nicht mehr exakt gleich $c$, sonder gleich $c\sqrt{1-\beta^2}$, aber immerhin bleibt sie hier isotrop, im ausgezeichneten System sogar streng gleich $c$. Um schließlich den MICHELSON-Versuch *ohne Längenkontraktion* zu erklären, genügt es dann anzunehmen, daß die Lichtgeschwindigkeit schwach abhängt vom Bewegungszustand der Lichtquelle, und zwar im ausgezeichneten System gemäß $c_{parallel} = c$, $c_{senkrecht} = c\sqrt{1-\beta^2+\beta^4}$, wobei sich die Bezeichnungen ‚parallel' und ‚senkrecht' auf die Bewegungsrichtung der Lichtquelle beziehen. Eine solche hypothetische Abhängigkeit der Lichtgeschwindigkeit vom Bewegungszustand der Lichtquelle wird unseres Erachtens nicht einmal unmittelbar widerlegt durch das diesbezüglich ansonsten sehr aussagekräftige CERN-Experiment von ALVÄGER *et. al.* [23], die bekanntlich gemessen haben, daß die Geschwindigkeit von γ-Strahlen etwa in Richtung der Emitter-Geschwindigkeit höchstens in einem Verhältnis von ca. $10^{-4}$ von der Lichtgeschwindigkeit $c$ abweicht. Auf die Realität der Längenkontraktion läßt sich in praktisch durchführbaren Experimenten demzufolge nur indirekt schließen[43]), und zwar aus einer durchgängigen und uneingeschränkten Gültigkeit der speziellen Relativitätstheorie in allen Inertialsystemen.

box@peter-ostermann.de

---

[42]) Im Unterschied zu den allseits bestens bekannten *kinematischen* Paradoxien werden die in der genannten Arbeit vorgestellten *dynamischen* Paradoxa – die z.B. auf Grundlage der (unzutreffenden) Voraussetzung starrer Körper die notwendige Existenz von Tachyonen imaginärer Ruhemasse beweisen (was eine Widerlegung der Voraussetzung darstellt) – bis heute offenbar weitgehend übersehen. Unsere diesbezügliche Arbeitshypothese lautet: *Eine konsequent durchgeführte relativistische Mechanik (die ausgedehnte mikroskopische Teilchen als stationäre Gebilde behandeln muß) wird sich am Ende als Quantenmechanik erweisen*. Es ist geplant, in einer eigenen Arbeit darauf näher einzugehen.

[43]) Zwar hat EINSTEIN [24] bereits in den Anfangsjahren gezeigt, daß und wie die Längenkontraktion prinzipiell meßbar ist unter der Voraussetzung, daß die spezielle Relativitätstheorie ohne jede Einschränkung zutrifft. Doch stellt dieses Gedankenexperiment natürlich keinen *unmittelbaren* experimentellen Nachweis dar. – In dem gleichen kurzen Beitrag hat EINSTEIN trotz des verwendeten Titels EHRENFESTs Arbeit zwar auch nicht zitiert, zu dem Sachverhalt aber immerhin bemerkt, „ *... was E h r e n f e s t in sehr hübscher Weise deutlich gemacht hat"*. Unseres Erachtens zeigt dies, daß es für EINSTEIN ganz selbstverständlich war, einfach *alle* der damals (1911) noch leicht überschaubaren relevanten Artikel zur Relativitätstheorie – und insbesondere zu dem hochaktuellen EHRENFEST'schen Paradoxon – als bekannt vorauszusetzen (s. auch Fußnote [33]).

## Literatur